\documentclass[aoas,preprint]{imsart}

\RequirePackage[OT1]{fontenc}
\RequirePackage{amsthm,amsmath}
\RequirePackage{natbib}
\RequirePackage{url}
\RequirePackage{bm}
\RequirePackage{lscape}
\usepackage{graphicx,psfrag,epsf}
\RequirePackage{xcolor}
\RequirePackage{geometry}
\usepackage{algorithm}
\usepackage{algorithmicx}
\usepackage[normalem]{ulem}

\startlocaldefs
\numberwithin{equation}{section}
\theoremstyle{plain}

\endlocaldefs

\begin{document}

\begin{frontmatter}
\title{Predicting competitions by combining conditional logistic regression and subjective Bayes: An Academy Awards case study}
\runtitle{Predicting competitions}

\begin{aug}
\author{\fnms{Christopher T.} \snm{Franck}\thanksref{t1}\ead[label=e1]{chfranck@vt.edu}}
\and
\author{\fnms{Christopher E.} \snm{Wilson}\thanksref{t2}\ead[label=e2]{chris.wilson@time.com}}

\runauthor{Franck and Wilson}

\affiliation{Virginia Tech\thanksmark{t1} and TIME\thanksmark{t2}}

\address{Christopher T. Franck\\
Department of Statistics, Virginia Tech\\
403E Hutcheson Hall (0439)\\
250 Drillfield Drive\\
Blacksburg, VA 24061\\
\printead{e1}\\
\and \\
Christopher E. Wilson\\
TIME, Washington, D.C. Bureau\\
1130 Connecticut Ave. NW, Ste. 900\\
Washington, DC 20036\\
\printead{e2}}

\address{ \\
 \\
 \\
 \\
 }
\end{aug}

\begin{abstract}
Predicting the outcome of elections, sporting events, entertainment awards, and other competitions has long captured the human imagination. Such prediction is growing in sophistication in these areas, especially in the rapidly growing field of data-driven journalism intended for a general audience as the availability of historical information rapidly balloons. Providing statistical methodology to probabilistically predict competition outcomes faces two main challenges.  First, a suitably general modeling approach is necessary to assign probabilities to competitors.  Second, the modeling framework must be able to accommodate expert opinion, which is usually available but difficult to fully encapsulate in typical data sets.  We overcome these challenges with a combined conditional logistic regression/subjective Bayes approach. To illustrate the method, we re-analyze data from a recent Time.com piece in which the authors attempted to predict the 2019 Best Picture Academy Award winner using standard logistic regression. Towards engaging and educating a broad readership, we discuss strategies to deploy the proposed method via an online application.

\end{abstract}


\end{frontmatter}

\section{Introduction}
\label{sec:intro}

Humans are naturally interested in competition. In the buildup to any contest of public interest, spectators have long made a study of predicting the outcome ahead-of-time, through a combination of instinct, prior observations, domain expertise, and, more recently, sophisticated analysis of data from past contests. Media outlets frequently capitalize on this interest by forecasting the winners of upcoming sporting events, entertainment awards, and elections, sometimes months in advance.

Until relatively recently, predictions published or broadcast to general-interest audiences were largely the product of seasoned domain experts who either relied entirely on personal insight or, if they incorporated data, rarely approached the task with statistical rigor. This sort of prognostication, particularly in politics, is so popular that, coterminous with the rise of 24-hour cable news stations, the word “pundit” took on a new coinage as a professional media forecaster \citep{Chertoff2012ironic}.

Despite recent interest by data experts in predicting competition outcomes, pundits dominated the popular market for some time.  As the information age has made relevant data across many domains more available without subscription to expensive proprietary databases, data-analytic approaches have been established for generating popular predictions for elections \citep{Rothschild2012obama, Linzer2013dynamic, Linzer2016votamatic, silver2012nov, Silver2016who}, sporting outcomes \citep{KerrDineen2017win, tango2007book, Nguyen2015making} and entertainment awards, \citep{King2019approximating}. Given the relatively recent influx of data analytic approaches, a tension can emerge between data-driven models and opinion-driven subjective punditry. For examples in politics, see \cite{Byers2012nate} and \cite{Cohn2017review}. For examples in sport, see \cite{KerrDineen2017win} and \cite{Lengel2018has}. The purpose of this work is to propose a modeling approach that probabilistically predicts the outcome of upcoming competitions by using both historical data and contemporary expert knowledge.

\begin{figure}[ht!] 
\centering
\includegraphics[width=100mm,trim=0 0 0 0]{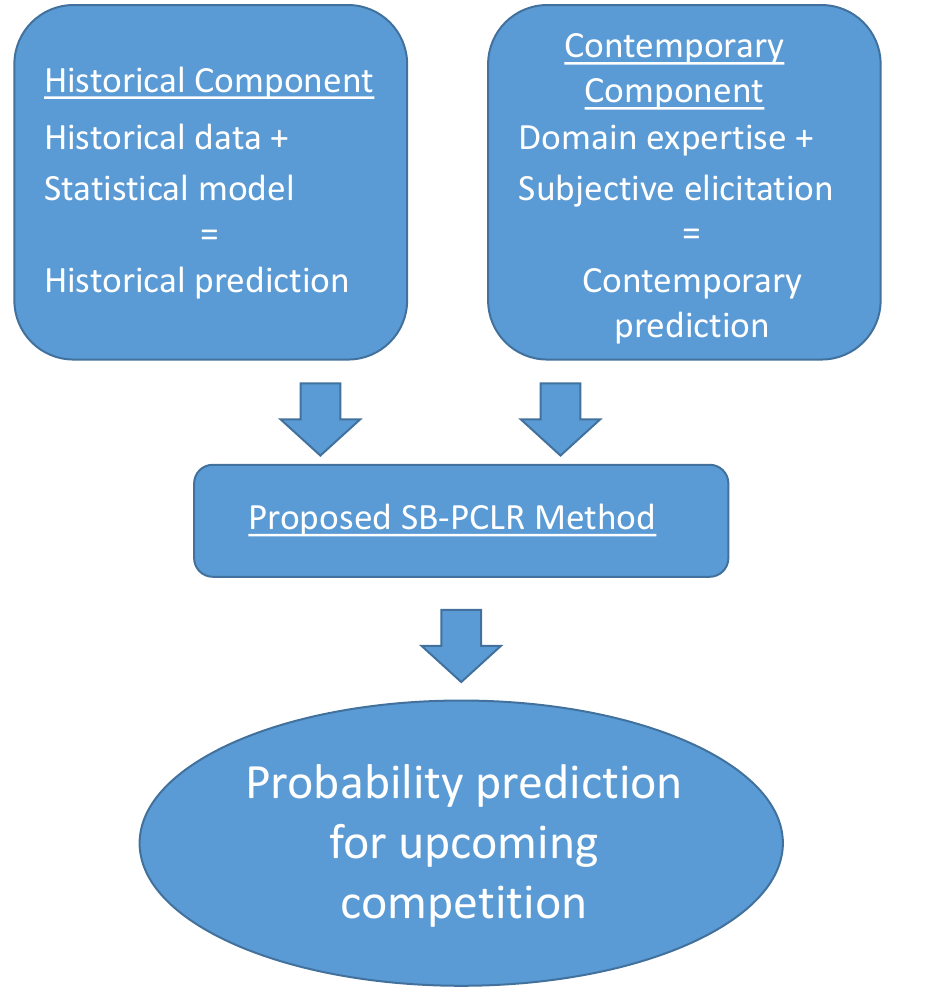}
\caption[]{Overview of the SB-PCLR} approach.
\label{fig:method_overview}
\end{figure}

Most reasonable people would agree that there is predictive value in historical competition data, but there are also usually important contemporary forces that are not trivial to represent in a historical data analysis. To illustrate this dichotomy, consider the upcoming case study in Section \ref{sec:casestudy} where we develop predictions for the winner of the 2019 Academy Award (aka ``Oscar'') for Best Picture using only data available before the 2019 award show. A savvy  analyst might gather data from the Directors Guild of America (DGA) awards, which occur before the Academy Awards are announced. This analyst would note the strong association between  win-status of Best Director from the DGA, and the eventual win-status of Academy Award for Best Picture.  Between 1950 and 2018, the odds ratio for these two variables is $66.6$ ($95\%$ bootstrap confidence interval: $27.6-168.9$). Owing to the strong statistical relationship, and since the DGA announces its winners before the Academy Awards, the DGA award for Best Director is a potentially useful candidate predictor for the Academy Award for Best Picture.

Historical predictors such as these are easy to incorporate into statistical models but do not address unique contemporary forces in a given round of the competition. For example, the members of the Academy who elected the winners in 2019 potentially felt resentment towards Netflix for disrupting the traditional Hollywood business model by funding, promoting, and controlling the distribution of their own content \citep{Brody2019my, Lawson2019oscar}. Since the 2019 Best Picture nominee \textit{Roma} was distributed and promoted by Netflix, a wise prognosticator would have weighed this contemporary knowledge heavily in their 2019 prediction. Our previous historical analysis suggested \textit{Roma} was the most likely candidate to win in 2019 \citep{Wilson2019we}, but the film \textit{Green Book} ultimately won. Thus, to account for both historical trends and also contemporary expert knowledge going forward, our proposed method incorporates subjective effects alongside the historical analysis as shown in Figure \ref{fig:method_overview}. We call our method the subjective Bayesian predictive conditional logistic regression (SB-PCLR) approach.

The SB-PCLR method we propose in Section \ref{sec:method} is flexible and applies to competitions (a) for which historical data is available, (b) that have an upcoming round with known contestants (e.g. the next football game or season, the next award show, the next election), and (c) the number of winners is fixed, where we focus on the single winner case. Since the identity of contestants in upcoming rounds of the competition is usually known in advance, we develop a conditional logistic regression approach for our predictions. Importantly, we develop the notion of ``prospective strata,'' which enables out-of-sample prediction for conditional logistic regression.  Traditional conditional logistic regression based on matching does not permit out-of-sample prediction.  See Section \ref{sec:method} for further detail on our strategy to extend conditional logistic regression for out-of-sample prediction. We allow for multiple-contestant competitions including award shows, division championships, and election primaries.

We take a Bayesian approach in this work, which enables natural incorporation of subjective effects into the analysis. Since this work is a data journalism-inspired effort, we describe strategies to implement the modeling approach in an interactive web-based format that should be appealing to a broad readership. We present Markov chain Monte Carlo techniques and we also develop a \textit{maximum a posteriori} (MAP) approach that is computationally faster for use in an online application targeted at a broad readership with a primary focus on point estimation.

We illustrate the SB-PCLR method by re-analyzing data from a recent Time.com piece \citep{Wilson2019we} which attempted to predict the 2019 Best Picture Academy Award winner using standard logistic regression. The new approach showcases an ability to model all future contestants on a probability scale with a sum-to-one constraint and also to include elicited subjective effects in the analysis.

In addition to the methodological SB-PCLR approach proposed here, we also compare and contrast the publication process between academic peer-reviewed journals and journalism venues.  Statisticians would appear to be well-poised to make meaningful contributions to the nascent area of analytic-powered journalism, yet such collaborations are not especially common.  From our team's experience, we briefly describe some challenges that face such collaborations and some suggestions to overcome  in Section \ref{sec:disc}. 

The remainder of this article is organized as follows.  Section \ref{sec:method} describes the proposed SB-PCLR method.  Section \ref{sec:casestudy} revisits the 2019 Academy Awards and illustrates our approach to predicting the winner of Best Picture using data from 1950-2018.  Section \ref{sec:disc} includes discussion of broader implications of the SB-PCLR  method, future plans and directions, and commentary pointed towards facilitating greater cooperation between statisticians and data journalists.

\section{The SB-PCLR method and comparators}
\label{sec:method}

Let $\bm{Y}$ be an $N \times 1$ vector of binary outcomes such that elements of $\bm{Y}$ are equal to one for winners and zero otherwise. Let $\bm{X}$ be a $N \times p$ model matrix that contains variables useful for predicting $\bm{Y}$.  Logistic regression is probably the most widely used approach to model $\bm{Y}$ as a function of $\bm{X}$ when observed data are available. However, we are interested in modeling historical competitions and predicting prospective outcomes in cases where each round has only one winner. The usual logistic regression approach ignores this ``single winner'' constraint, and thus predicted probabilities of winning are formed in terms of the entire historical data and do not sum to one within any given round of the competition. The authors used this standard approach in a previous analysis of nominees for the 2019 Academy Award for Best Picture \citep{Wilson2019we}.

\subsection{Conditional logistic regression for out-of-sample prediction}
\label{subsec:condlogitreg}

For the purpose of estimating win probabilities, a more satisfying approach is to constrain predicted probabilities to sum to one within each round. For example, if we wish to predict the winner of the Academy Award for Best Picture in a future year, we would like the finalists' predicted probabilities to sum to one so that each probability represents a chance of winning the upcoming contest relative to the specific participants in that round of the contest. We impose the sum-to-one constraint by developing a conditional logistic regression approach that enables inference on historical effects and exploits the known structure of upcoming competitions to enable out-of-sample prediction. 

Let $k=1,\ldots,K$ index the strata, and let $\bm{Y}^{(k)}$ and $\bm{X}^{(k)}$ represent the subset of the outcome vector and model matrix corresponding to the $k$-th stratum, i.e. 

\[
\bm{Y}
=
\begin{pmatrix}
    \bm{Y}^{(1)} \\
    \vdots \\
    \bm{Y}^{(K)}
\end{pmatrix}
\text{ and }
\bm{X}
=
\begin{bmatrix}
    \bm{X}^{(1)} \\
    \vdots \\
    \bm{X}^{(K)}
\end{bmatrix}.
\]
In the Academy Awards example, each year is a stratum $k$, $\bm{X^{(k)}}$ might include summaries of critical reception, commercial success, and other film-specific accolades. $\bm{Y^{(k)}}$ is then a length $n_k$ vector that contains a single one corresponding to the winner and zeroes elsewhere, i.e. $\sum_{i=1}^{n_k}y_i^{(k)}=1$ where $y_i^{(k)}$ is the outcome for the $i$-th competitor $i=1,\ldots,n_k$ in stratum $k$. 

The conditional likelihood function arises from conditioning on a fixed number of events (one in this work) per stratum. We closely follow the notation in \cite{Agresti2013categorical}. To establish the conditional likelihood, consider the set $S(1)$ of all possible $\bm{Y}$ vectors that have one and only one event. Define the set of such vectors as 

$$S(1) =\Big\{ \big(y_1^{*(k)},\ldots,y_{n_k}^{*(k)} \big):\sum_{i=1}^{n_k} y_i^{*(k)} =1 \Big\},$$

\noindent
where $y_i^{*(k)}$  refers to the possible observed outcomes such that each stratum has one and only one event. By the rules of conditional probability, the user can (a) plug each set of $y_i^{*(k)}$ values that obey the sum-to-one constraint into the logistic regression likelihood, (b) sum this likelihood over the $S(1)$ set, and (c) divide the usual logistic regression likelihood by this sum to obtain the conditional logistic regression likelihood. The conditional likelihood for the $k$-th stratum is thus

\begin{align}
\label{stratalike}
P\Big(\bm{y}^{(k)}|\bm{\beta}\Big) =P\Big(Y^{(k)}_1=y^{(k)}_1,\dots,Y^{(k)}_{n_k}=y^{(k)}_{n_k}| \sum_{i=1}^{n_k} y_i^{(k)}=1, \bm{\beta}\Big) \\
 = \frac{\text{exp}\Big[\sum_{j=1}^p\big(\sum_{i=1}^{n_k}y_i^{(k)}x_{ij}^{(k)}\big)\beta_j\Big]}{\sum_{S(1)}\text{exp}\Big[\sum_{j=1}^p\big(\sum_{i=1}^{n_k}y_i^{*(k)}x_{ij}^{(k)}\big)\beta_j\Big]}, \nonumber
\end{align}

\noindent
where $x_{ij}^{(k)}$ is the $j$-th predictor variable for the $i$-th competitor, $\beta_j$ is the coefficient for the $j$-th predictor, and $j=1,\ldots,p$. Given the regression effects $\bm{\beta}$, we assume independence between strata. Thus the conditional likelihood function is the product of the strata-level conditional likelihood functions

\begin{equation}
P\Big(\bm{Y}^{(1)}=\bm{y}^{(1)},\dots,\bm{Y}^{(K)}=\bm{y}^{(K)}|\sum_{i=1}^{n_k} y_i^{(k)}=1\text{ for }k=1,\ldots,K,\bm{\beta}\Big)=\prod_{k=1}^K P\big(\bm{y}^{(k)}|\bm{\beta}\big).
\label{histlike}
\end{equation}

Conditional logistic regression was developed in 1978 \citep{Breslow1978estimation}. A review of citations in this area reveals that the technique has been popular to address stratification in the fields of medicine and epidemology. Conditional logistic regression forms strata by matching data points post-data collection based on similar characteristics such that there is a fixed number of events occurring per stratum. A canonical example would involve forming strata at the patient-level based on similarities among patients, such that each strata has a fixed number of events (e.g. illness, death) per stratum. Since patients are placed into strata via matching post-data collection, prospective out-of-sample prediction is unavailable because there is no natural concept of a strata for patients who were not matched within the study. For example, out-of-sample patients do not belong to strata for which the number of events is fixed and known, which makes it impossible to directly adapt Equation (\ref{stratalike}) when forming a likelihood. There is no way to condition on a fixed number of events per strata when incoming patients do not appear in the context of strata. This appears to be the reason for the widespread perception that conditional logistic regression is not available for out-of-sample prediction. For a further review of conditional logistic regression, see \cite{Agresti2013categorical} 

The SB-PCLR method takes a different approach by forming strata composed of the competitors in an upcoming round of a competition. Fortunately, for the competitions we describe in this work, the identities of competitors for upcoming competitions are known before the competition takes place. Therefore, useful predictor variables can be gathered and the sum-to-one constraint can be imposed on the likelihood for upcoming data. We thus develop the notion of a prospective strata and incorporate its likelihood into (\ref{histlike}) to enable out-of-sample probability prediction for the winner of the upcoming contest. 

Since the main goal of this work is to probabilistically predict future outcomes, we next formally develop the idea of a prospective stratum $C$. Let $C$ represent the $(K+1)$-th stratum and $n_C$  represent the number of competitors in this stratum. For example, in Section \ref{sec:casestudy} the $C$-th stratum is comprised of the films which were nominated for the 2019 Academy Award for Best Picture. We know that only one film will win, and the predictors $\bm{X}^{(C)}$ were observed between the announcement of nominees in late January and the award show in February of 2019. In cases such as these, the $C$-th stratum likelihood can be formed as:

\begin{align}
\label{newstratalike}
P\big(\bm{y}^{(C)}|\bm{\beta}\big) =P\Big(Y^{(C)}_1=y^{(C)}_1,\dots,Y^{(C)}_{n_C}=y^{(C)}_{n_C}| \sum_{i=1}^{n_C} y_i^{(C)}=1, \bm{\beta}\Big) \\
 = \frac{\text{exp}\Big[\sum_{j=1}^p\big(\sum_{i=1}^{n_C}y_i^{(C)}x_{ij}^{(C)}\big)\beta_j\Big]}{\sum_{S(1)}\text{exp}\Big[\sum_{j=1}^p\big(\sum_{i=1}^{n_C}y_i^{*(C)}x_{ij}^{(C)}\big)\beta_j\Big]}. \nonumber
\end{align}

Of course, the outcome data for the upcoming round of the competition represented by $\bm{y}^{(C)}$ is not available at the time of inference. Thus, inferences on $\bm{\beta}$ are based on the conditional likelihood (\ref{histlike}). Probabilistic predictions for $\bm{y}^{(C)}$ are formed using estimates of $\bm{\beta}$ and the $\bm{X}^{(C)}$ predictors using the prospective strata likelihood (\ref{newstratalike}). We take a Bayesian approach to inference and prediction in this paper with an eye towards incorporating subjective effects in Subsection \ref{subsec:subjeff}. For the case study presented in Section \ref{sec:casestudy}, we include both MAP estimates and Markov Chain Monte Carlo (MCMC) results from a Metropolis sampler to conduct statistical inference and make probabilistic predictions using the model chosen in \cite{Wilson2019we}.

\subsection{Incorporating subjective effects}
\label{subsec:subjeff}

The final aspect of the SB-PCLR method involves optional specification of subjective effects for incorporation into the analysis. When upcoming competitions capture the public's interest, enthusiasts, pundits, and prognosticators may wish to incorporate their opinions about the individual competitors in the prospective $C$-th stratum into the analysis formally. Let $Q=n_C-1$, and let $\bm{\phi}=\langle\phi_1,\ldots,\phi_Q\rangle^T$ represent subjective effects on the logit scale specific to the $n_C$ competitors in the prospective $C$-th stratum. Our model is parameterized using $Q$ competitor-specific effects which represent the change in log odds of winning associated with moving from a baseline competitor to the $q$-th competitor, $q=1,\ldots,Q$. This parameterization is similar to the baseline category parameterization for multinomial logistic regression, see e.g. \cite{Agresti2013categorical}. While it is possible to incorporate subjective effects into any strata the user wishes, we focus solely on subjective effects for the prospective $C$-th stratum, i.e. the upcoming competition.

Eliciting subjective effects on a relative log odds scale is not intuitive for most people. Probability scales are convenient for subjective elicitation \citep{Ohagan2006uncertain}, so we recommend asking users to provide subjective win probabilities for each prospective competitor such that these probabilities sum to one. Let $p_c$ represent the $c$-th prior subjective probability such that $0 \leq p_c \leq 1$ and $\sum_{c=1}^{n_C} p_c=1$ for $c=1,\ldots,n_C$.  To obtain $\bm{\phi}$ from these user-specified win probabilities, consider the film $p_{n_C}$ as the baseline, 

$$\phi_q = log\Big(\frac{p_q}{p_{n_C}}\Big) \text{ for } q=1,\ldots,Q,  $$

\noindent
and $\phi_q$ measures the shift in log-odds of winning between the baseline competitor and the $q$-th competitor, $q=1,\ldots,Q$.

We incorporate subjective effects into the analysis via a mixture model approach. Let

\begin{equation}
\label{mixture}
P\big(\bm{y}^{(C)}|\bm{\beta},\bm{\phi},\omega\big) = \omega P\big(\bm{y}^{(C)}|\bm{\beta}\big) + (1-\omega)P\big(\bm{y}^{(C)}|\bm{\phi}\big),
\end{equation}

\noindent
where $\omega$ is a mixing weight that governs how heavily the historical model effects $\bm{\beta}$ are weighed relative to the subjective effects $\bm{\phi}$, and  

\begin{equation}
\label{}
P\big(\bm{y}^{(C)}|\bm{\phi}\big)= \frac{\text{exp}\Big[\sum_{q=1}^Q\big(\sum_{i=1}^{n_C}y_i^{(C)}z_{iq}^{(C)}\big)\phi_q\Big]}{\sum_{S(1)}\text{exp}\Big[\sum_{q=1}^Q\big(\sum_{i=1}^{n_C}y_i^{*(C)}z_{iq}^{(C)}\big)\phi_q\Big]}\
\end{equation}
is essentially a $C$-th stratum conditional likelihood function based on subjective effects $\bm{\phi}$. The indicator variable $z_{iq}^{(C)}=1$ if $q=i$ and zero otherwise. This indicator's role is to apply the correct film-specific subjective effect to each film relative to baseline. Thus, for the prospective $C$-th stratum, we model win probability as a function of both the historical prediction mixture component and the subjective component.  We envision users of this approach will vary in the extent to which they rely on the historical model versus their own subjective opinion of the competitors.

Finally, let the complete data vector $\bm{y}^{(full)}=\langle \bm{y},\bm{y}^{(C)}  \rangle^T$, and

\begin{equation}
\label{fullmod}
P\big(\bm{y}^{\text{(full)}}|\bm{\beta},\bm{\phi},\omega\big) = P\big(\bm{y}^{(C)}|\bm{\beta},\bm{\phi},\omega\big) \times \prod_{k=1}^K P\big(\bm{y}^{(k)}|\bm{\beta}\big).
\end{equation}

The parameters in full likelihood (\ref{fullmod}) are $\bm{\beta}$, $\bm{\phi}$, and $\omega$. Our analysis proceeds in a Bayesian fashion, where $\bm{\beta}\sim N(\bm{0},\sigma^2 I)$ is given a vague proper prior and $\omega$ and $\bm{\phi}$ are subjectively elicited from the audience interested in the outcome of the competition.  In practice, the $\bm{y}^{(c)}$ data are not yet observed and thus the likelihood in Equation (\ref{fullmod}) cannot directly be used for inference on future films. Instead, we obtain posterior summaries of the $\bm{\beta}$ parameters using MAP or MCMC methods, solicit $\bm{\phi}$ values from domain experts, then use these values alongside the $\bm{X}^{(C)}$ predictor values to estimate the probability that each prospective competitor wins the upcoming $C$-th round of the competition. 

We briefly discuss the challenge in eliciting suitable variability for $\omega$ and $\bm{\phi}$ here and include a more thorough discussion in Section \ref{sec:disc}. Conceptually, there is no technical difficulty imposing prior distributions on $\omega$ and $\bm{\phi}$, where the Beta and Normal families of distributions seem like a reasonable starting point, respectively.  Assessing variability in subjective opinion for the SB-PCLR model is an unresolved issue.  If prior precision on the $\bm{\phi}$ effects is low, the subjective mixture component will influence the posterior towards uniform probabilities for the competitors. Since we anticipate deploying this model via a web interface among readers who are unfamiliar with the nuances of Bayesian analysis and prior specification, we wish to restrict the required inputs to $\omega$ and  $p_c$ for $c=1,\ldots,n_C$. We anticipate that users who do not follow the subject matter or are not confident in their personal expertise would reduce the $\omega$ value towards zero to reduce the impact of their choice of $\bm{\phi}$ , while those who trust their instincts over historical trends, or believe the current contest is radically different from all prior ones, would place the $\omega$ value at or near 1 and carefully adjust the $\bm{\phi}$ terms. If the user's goal is to favor uniform outcomes among films, we anticipate they would prefer to specify a near-uniform specification of their prior win-probabilities $p_c$ for $c=1,\ldots,n_C$ rather than specifying low precision in the $\bm{\phi}$ terms.

Since we suggest deploying our method to a broad readership, it is possible to learn about the distribution of parameters $\omega$ and $\bm{\phi}$ across many subjects empirically. We deployed a strategy like this in our followup article \citep{Wilson2020our} to the original Academy Awards data journalism piece \citep{Wilson2019we}. To highlight how such an approach can work, we have developed a hand-built custom JavaScript online application that allows users to specify $\omega$ and $\bm{\phi}$ (via $p_c$), then receive probability predictions for the films under consideration in this case study. The application may be accessed here https://mech.sc/oscars. To ensure that users do not suggest greater than 100\% subjective probability, our application features an interesting reservoir system on the sliders. As the user increases subjective probability on a given film winning, the upper bound on remaining sliders decreases commensurately. We don't allow the user to submit subjective probabilities until the reservoir is empty, and we include the option of apportioning the remaining probability uniformly among films. Studying variability in responses from on this application may inform choices about prior specification in future studies.

The approach we implemented follows:
\begin{itemize}
\item[1.] At the participant level, treat user-specified values of $\omega$ and $\bm{\phi}$ as fixed, known constants for the competition under consideration.
\item[2.] Using an Institutional Review Board (IRB)-approved protocol, gather empirical data on selected values of $\omega$ and $\bm{\phi}$ for all consenting users \citep[TIME coauthor Wilson has implemented this strategy previously,][]{Wilson2019how}.
\item[3.] As a future research endeavor, the empirical distributions of $\omega$ and $\bm{\phi}$ can be used to inform plausible values of variability in future studies which use this approach.
\end{itemize}

In Section \ref{sec:casestudy} we use a standard Metropolis sampler to obtain point estimates and credible intervals for win probabilities for various specifications of $\omega$ and $\bm{\phi}$. Since the SB-PCLR method has been and likely will continue to be used via interactive web applications embedded in general interest stories, we compare the Metropolis sampler with a plug-in strategy based on MAP estimates, where the latter strategy reduces computation times. Readers can then have the opportunity to explore competition prediction in an interactive Bayesian environment in real-time. 

\subsection{Inferential process}
\label{subsec:infproc}

To explore the posterior distributions for the $\bm{\beta}$ coefficients and $P\big(\bm{y}^{(C)}|\bm{\beta}\big)$ values, we implement a Metropolis sampler. Algorithm \ref{alg:MetroMCMC} outlines the approach. Note that in line 5, $P\big(\bm{\beta}|\bm{y}\big)$ refers to the posterior distribution of the $\bm{\beta}$ coefficients based on the conditional logistic regression likelihood (\ref{histlike}), $\bm{\beta}^{(prop)}$ represents the vector of coefficients where the $j$-th element contains the proposed values $\beta_j^{*}$ and all other elements are current values, and $\bm{\beta}^{(\text{cur})}$ represents the current value of all coefficients. This algorithm may be initialized using MAP estimate values.
\begin{algorithm}
\caption{Metropolis sampler}
\label{alg:MetroMCMC}
\begin{algorithmic}
\State Initialize $\bm{\beta}^{(0)} = (\beta_1^{(0)},\ldots,\beta_p^{(0)})^T$.
\State For $t$ in $1$ to $T$ \{
\State \hskip1.0em  For $j$ in $1$ to $p$ \{
\State \hskip2.0em Propose a new value for $\beta_j^{(prop)}$ from $N(\beta_j^{(t-1)},\tau^2)$.
\State \hskip2.0em Calculate acceptance ratio $\alpha = P\big(\bm{\beta}^{(prop)}|\bm{y}\big)/P\big(\bm{\beta}^{(\text{cur)}}|\bm{y}\big)$.
\State \hskip2.0em Sample $u \sim \text{Uniform}(0,1)$. If $u<\alpha$ then set $\beta_{j}^{(t)}= \beta_j^{(prop)}$; else let $\beta_j^{(t)}=\beta_j^{(t-1)}$.
\State \hskip1.0em \}
\State \hskip1.0em Compute historical $P\big(\bm{y}^{(C)}|\bm{\beta}^{(t)}\big)^{(t)}$ using Equation (\ref{newstratalike}) for all possible values of $\bm{y}^{(C)}$.
\State \hskip1.0em Compute subjectively-informed $P\big(\bm{y}^{(C)}|\bm{\beta},\bm{\phi},\omega\big)$ using Equation (\ref{mixture}) with user-provided $\bm{\phi}$ and $\omega$ for \State \hskip1.4em all possible values of $\bm{y}^{(C)}$.
\State \}
\end{algorithmic}
\end{algorithm}

\subsection{Approximate Bayesian model selection on historical predictors}
\label{subsec:ABMA}

A downside of the chosen model in our original Time.com \citep{Wilson2019we} article is that several variables which are positively associated with winning Best Picture are not considered when making probabilistic predictions based on historical data. Bayesian model averaging is a satisfying approach that incorporates uncertainty due to unknown model structure because it uses posterior probability of a model being correct to provide weights for quantities of interest across a list of candidate models. It seems likely that some other good models exist, and perhaps we should weight our predictions in a way that incorporates those models. To get a sense of whether model averaging would incorporate additional predictors with tangible weight and potentially alter our historical win probabilities, we implemented approximate Bayesian model averaging (ABMA) using the Bayesian Information Criterion \citep{Schwarz1978estimating}.  

Bayesian model averaging relies on the posterior probability of each candidate model being correct given the observed data. Let $m=1,\ldots,M$ index the $M=1023$ models formed by considering every subset of the top ten predictors listed in Section \ref{subsec:casestudyresults}. (Note that for conditional logistic regression, there are no intercept terms and thus no intercept-only model.) The posterior probability of model $M_m$ given observed data $\bm{y}$ is

\begin{equation}
P(M_m|\bm{y})=\frac{P(\bm{y}|M_m)P(M_m)}{\sum_{r=1}^R P(\bm{y}|M_r)P(M_r)}, 
\label{eq:postprob}
\end{equation}

\noindent
where $P(M_m)$ is the prior probability on model $M_m$ and $P(\bm{y}|M_m)$ is marginal distribution of the data $\bm{y}$ given model $M_m$, which arises from integrating the parameters out of the joint distribution of the data and parameters.

The familiar Bayesian Information Criterion \cite[BIC,][]{Schwarz1978estimating} can be used to asymptotically approximate posterior model probabilities \citep{Kass1995bayes} in Equation (\ref{eq:postprob}) as follows. As n increases, the quantity

\begin{equation}
e^{-1/2(BIC_i-BIC_j)} \approx \frac{P(\bm{y}|M_i)}{P(\bm{y}|M_j)}
\label{eq:BICapprox}
\end{equation}

\noindent
where the quantity on the right hand side of Equation (\ref{eq:BICapprox}) is a Bayes Factor, which summarizes the multiplicative change from prior to posterior odds of two competing models $M_i$ and $M_j$ for observed data $\bm{y}$ \citep{Kass1995bayes}. Then, for specified model prior probabilities, some algebraic manipulation can be performed on the right hand side of Equation (\ref{eq:BICapprox}) to approximate the posterior model probabilities $P(M_m|\bm{y})$ in Equation (\ref{eq:postprob}).

Bayesian model averaging proceeds by weighting the posterior probability distributions of quantities of interest across all candidate models proportionally to each models' posterior probability \citep{Hoeting1999bayesian}. Let $\Delta$ represent a quantity of interest that is common to all candidate models. In this study we are interested in model averaging regression coefficients and predicted win probabilities in the upcoming contest. Through the Law of Total Probability $P(\Delta|\bm{y})=\sum_{r=1}^R P(\Delta|M_r,\bm{y})P(M_r|\bm{y})$, where the candidate set of models is implicitly presumed to (a) contain the true model, and (b) form a partition of the model space. These assumptions are not necessarily exactly true, see Section \ref{sec:disc} for more discussion. Nonetheless, Bayesian model averaging is useful due to its ability to combine inferences across several plausible models, which helps to account for uncertainty due to model choice. For our approximate version in this paper, we substitute each models' point estimate in place of $P(\Delta|M_r,\bm{y})$ and BIC-approximated posterior model probabilities for $P(M_r|\bm{y})$. 

\subsection{Regularization-based approach to prediction}
\label{subsec:LASSO}

In order to draw additional comparisons between SB-PCLR methods and other available techniques, we used the \textit{glmnet} package \citep{Friedman2010regularization} in R to implement a regularization-based approach to the Academy Awards case study in Section \ref{sec:casestudy}.  To implement the approach, we used the least absolute shrinkage and selection operator (LASSO) \textit{l1} norm (rather than a blend of LASSO and ridge). We used k-fold cross validation to set the tuning parameter via the \textit{cv.out()} function, and we used the pre-2019 data for training to make predictions on the 2019 contest. In an attempt to avoid overfitting, we used the tuning parameter within one cross validation standard error of the lowest mean square error tuning parameter, as is relatively standard. Unlike the two-stage approach where we identified the top 10 predictors for further selection, we used all 47 candidate predictors in the LASSO approach. This off-the-shelf approach to prediction does not account for stratification, constrain the win probabilities in a year to sum to one, or naturally accommodate subjective effects specification.

\section{Case study: 2019 Academy Award for Best Picture}
\label{sec:casestudy}

The Academy of Motion Picture Arts and Sciences is a semi-secretive body of film professionals that annually issues awards for meritorious films. The Academy Awards, more popularly known as the Oscars, are given out in many categories, from technical categories like ``sound editing'' to highly coveted honors for the Best Actor and Actress and Best Director, culminating in Best Picture, for which up to ten films can be nominated (expanded from five in recent years). All members of the Academy are eligible to vote for Best Picture, and this award relies on a preferential voting system in which members rank the nominees in order of preference. This preferential voting system was introduced after 2009; prior to this the Academy issued awards based on whichever film captured the plurality of votes. Note the SB-PCLR implementation in this case study does not accommodate information related to the preferential data, as it is not publicly available. More details about the structure and voting for the Academy Awards can be found here \citep{Gray2020how}.

As mentioned in Section \ref{sec:intro}, this case study focuses on predicting the winner of the 2019 Academy Award for Best Picture using historical data from 1950-2018 and considering various choices of $\omega$ and $\bm{\phi}$. This case study is a re-analysis of data which were previously used to make probabilistic predictions using standard logistic regression with no subjective effects \citep{Wilson2019we}.

The eight films under consideration for Best Picture in 2019 were \textit{A Star is Born},  \textit{Black Panther}, \textit{BlacKKKlansman}, \textit{Bohemian Rhapsody}, \textit{Green Book}, \textit{Roma}, \textit{The Favourite}, and \textit{Vice}. For this analysis, we rely on the predictors we used previously \citep{Wilson2019we} so that we may compare our SB-PCLR approach to the standard unconditional logistic regression analysis used previously. For the SB-PCLR approach, we perform analysis via MAP estimation in the context of an online application, and we use a Metropolis sampler to provide credible intervals for inferences on $\bm{\beta}$ and probability predictions. We compared these approaches to ABMA and LASSO approaches described in Sections \ref{subsec:ABMA} and \ref{subsec:LASSO}.

For this exercise, we gathered historical data on all Oscar nominees for Best Picture since 1950. The first Academy Awards ceremony was held in 1929, but the earliest year where data are consistently available is 1950. Naturally, if the window of historical data is adjusted, the results of the SB-PCLR analysis could potentially change.   Nominees are announced in late January each year and awards are announced towards the end of February. The DGA, which is unaffiliated with the Academy, announces its nominees in early January and declares winners in late January. Thus, the Academy Award nominations for Best Picture candidates in other categories, like whether a potential winner also generated a Best Actress nomination, as well as the DGA nominations and winners can be used as candidate predictors of winning Best Picture in the same year. The nominees and winners for all relevant awards can be easily harvested from the official Web sites for the awards and fact-checked against sites like the Internet Movie Database for any possible discrepancy. While some of the lesser known technical awards have changed names and precise definitions since 1950, the data set is remarkably consistent across 59 years. We considered 47 possible binary input variables related to award nominations and wins, including a list of other Academy Awards for which the film was also nominated in the same year and a small list of awards given by other organizations, like the Directors Guild, that have consistently announced awards before the Academy’s. We note that Academy has historically considered five nominees per year, but as of 2010 the number of nominees varies between eight and ten.

To collect the data, we gathered a candidate list of potential variables by examining the pages for every past Best Picture winner on the Internet Movie Database (imdb.com), which includes a comprehensive list of nominations and wins for everything from the Academy Awards to the Golden Globes, the British Academy of Film and Television Arts (BAFTA), the Directors Guild of America, all the way down to the Dallas-Fort Worth Film Critics Association. Only those societies who nominated (and sometimes awarded) films before the Academy Awards ceremony in a consistent manner back at least to 1950 were considered, which reduced the variables to nominations for other Academy Awards in the same year; the Golden Globes; the BAFTAs; and the Directors Guild of America, all of which began in the 1940s. In some cases, like the Golden Globes, the winners are announced prior to the Oscars Ceremony, so each award provided two variables: whether a film was nominated and whether it won.

The data on each film’s nomination and win status, when relevant, was gathered from each award  official website and spot-checked against both the Internet Movie Database and the Open Movie Database API (http://www.omdbapi.com ). There was no evidence of disagreement in the historical data. Only awards that have been consistently granted for the same qualifications since 1950 were considered, which eliminated some recognitions of technical achievement that were not yet invented in 1950.

The original 2019 analysis was simplistic, since we used two-stage model selection based on BIC \citep{Schwarz1978estimating} and the standard (unconditional) logistic regression framework. The simplicity of our original model selection approach was motivated in part by the editorial deadline. We needed to publish our findings with enough lead time that our article would be of interest to the readership of Time.com.  Compared with academic statistics journals, the editorial timeline in journalism is much faster, which imposes constraints on the scope of analyses. See Section \ref{sec:disc} for more discussion on this issue. Issues related to model selection and averaging in the context of competition prediction are also discussed further in Section \ref{sec:disc}.  

In the first stage of the variable selection, we identified those predictors with the top 10 highest-magnitude odds ratios with the outcome. In the second stage, we formed an exhaustive list of the $2^{10}-1=1023$ models corresponding to all possible models with these candidates (the intercept term cancels in the conditional likelihood, so there is no intercept-only model). The model with the lowest BIC was selected. We did not consider interaction effects in this study.  This two-stage BIC approach ultimately selected three binary variables which we denote $x_1,x_2,x_3$:

\begin{itemize}
\item[$x_1$:] Whether the film was also nominated for the Academy Award for Best Director
\item[$x_2$:] Whether the film was also nominated for the Academy Award for Best Editing
\item[$x_3$:] Whether the film won top honors from the Directors Guild of America, which is announced before the Academy Awards
\end{itemize}

The user's choice of $\omega$ and $\bm{\phi}$ governs the extent to which the historical model component is weighed against subjective opinion in the production of win probabilities. In this study, we considered three specific prior settings for subjective probabilities. The ``GB prior'' assigns \textit{Green Book} (the eventual winner) 80\% of the prior probability and splits the remainder among the nine other candidates.  The ``U prior'' assigns each of the eight films $12.5\%$ of the prior probability of winning.  The ``NR prior'' reflects a disposition of someone who thinks \textit{Roma} is unlikely to win (perhaps due to Netflix's role producing the film), where \textit{Roma} has a one percent prior probability of winning and each other film has a uniform share of the remainder.

\subsection{Case study results}
\label{subsec:casestudyresults}

Tables \ref{Table1} and \ref{Table2} summarize predictions for the eight films under consideration. Table \ref{Table1} includes predictions from unconditional logistic regression, LASSO, the fully historical model (i.e. $\omega=1.0$) using SB-PCLR with predictors $x_1$, $x_2$, and $x_3$, three prior specifications for subjective effects at various $\omega$ values, and point estimates via ABMA. Table \ref{Table2} provides MCMC point estimates and credible intervals via the Metropolis sampler in Algorithm 1 for probability predictions for the historical model and three subjective effect specifications at a variety of $\omega$ values. Table \ref{Table3} shows the top ten predictors as identified by odds ratio, the observed odds ratios, univariate and historical model MCMC output for$\beta_j$ coefficients, LASSO, and ABMA results. As mentioned previously, the LASSO approach considers all 47 candidate variables, and the only non-zero LASSO coefficients observed corresponded to variables that had high odds ratios and thus are included in Table \ref{Table3}. Therefore, every non-zero LASSO coefficient is presented in Table \ref{Table3}.

\begin{figure}[ht!] 
\centering
\includegraphics[width=150mm,trim=0 0 0 0]{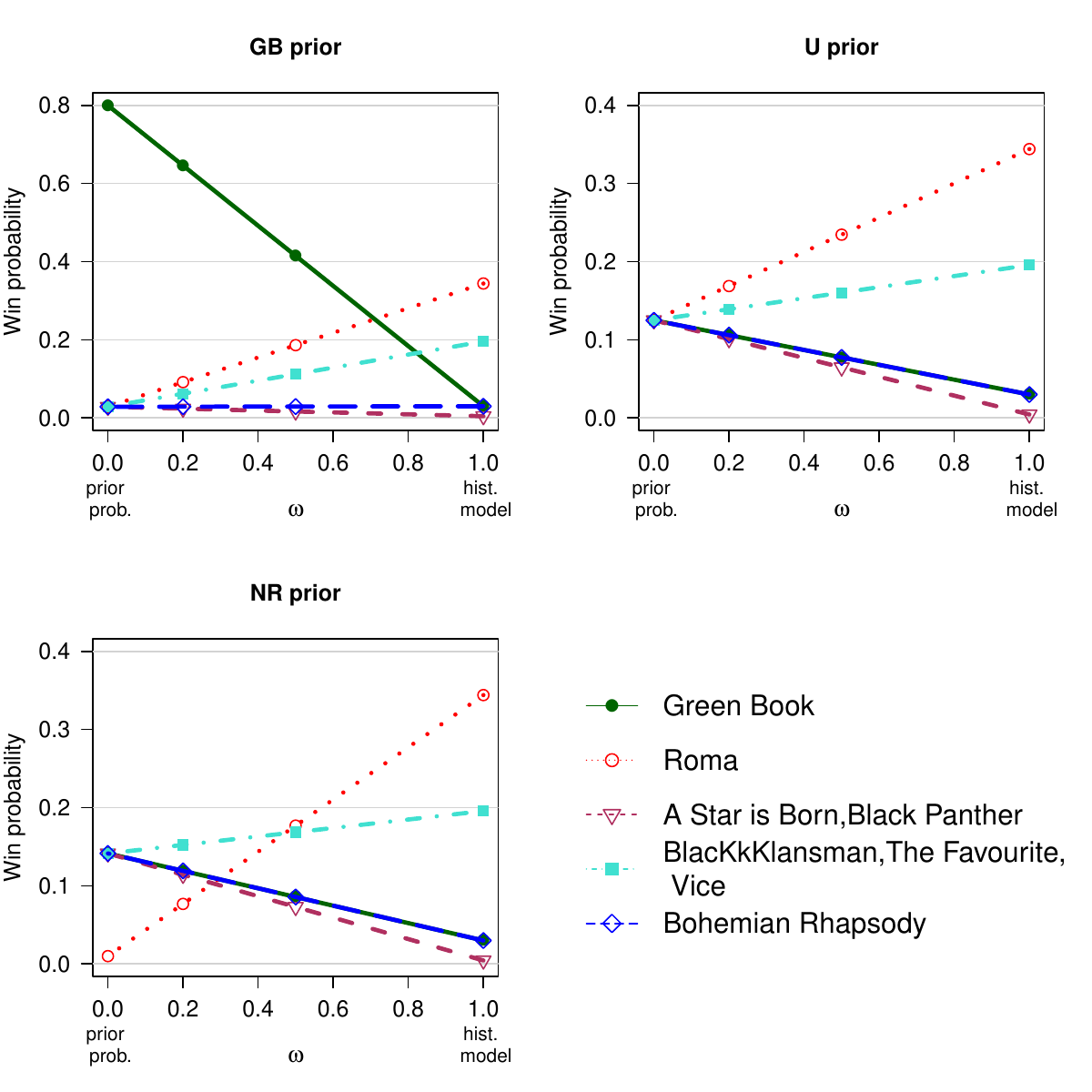}
\caption[]{Line plots for the case study analysis.}
\label{fig:results}
\end{figure}

Figure \ref{fig:results} shows the posterior predicted probabilities for three specific choices of $\bm{\phi}$ for various $\omega$ with the ``GB prior''  in the top left, ``U prior'' in the top right, and the ``NR prior'' in the bottom left. The left side of each horizontal axis corresponds to the prior probabilities in each of the ``GB,'' ``U,'' and ``NR'' priors.  Moving from left to right corresponds to an increase in $\omega$ and hence more weight on the historical model. At $\omega=1$, the prior probabilities on films are completely outweighed by the historical model. Figure \ref{fig:results} was constructed using the MAP strategy described in Section \ref{sec:method}.  Table \ref{Table1} includes probabilities shown via characters in Figure \ref{fig:results}. 

In addition to showing how various prior choices affect win probabilities, the line plots in Figure \ref{fig:results} are useful for ``\textit{post-mortem}-style'' analyses. A natural question is: \textit{How strongly would a prognosticator have needed to believe in Green Book (the eventual winner) in order to conclude that it had a better chance of winning than Roma?} The SB-PCLR modeling framework reveals that a prognosticator who valued the historical model at $\omega=0.7$ would have needed 80\% prior probability favoring \textit{Green Book} in order to conclude that the eventual winner was a more likely candidate than \textit{Roma}, which captures the plurality of posterior probability based on this particular historical model. This statement is reflected in the top left panel of Figure \ref{fig:results}. The astute reader will notice that in the ``U'' and ``NR'' settings, \textit{Green Book} and \textit{Bohemian Rhapsody} are tied.  This is because they share equal prior weight and happen to also share the same values for $x_1$, $x_2$, and $x_3$.

The ABMA results in Table \ref{Table1} summarize how approximate model averaging based on historical data fares with this case study. \textit{Roma} did slightly better with 40\% probability under ABMA. Interestingly, the eventual winner \textit{Green Book} also outperformed the prediction based on the 2019 model. This is likely because models that included predictors besides $x_1$, $x_2$, and $x_3$ favored \textit{Green Book}, those models had positive posterior probability, and those effects are ignored in the 2019 historical model with $\omega=1.0$. Table \ref{Table3} shows model averaged coefficients, many of which are much closer to zero compared with the 2019 historical model.

We compare the LASSO to the other methods of prediction in Table \ref{Table1} and provide the LASSO coefficients in Table \ref{Table3}. At the cross-validated tuning parameter chosen for this study, only three predictors had non-zero coefficients, and all of these belonged to the set of top ten predictors based on odds ratio used for the other selection approaches in this paper. Table 1 shows that the LASSO approach gave the majority of win probability to \textit{Roma}, with all other competitors ranging between seven and nine percent. Since the regularization approach does not rely on stratification, the win probabilities do not sum to one (as is also the case with the unconditional logistic regression approach). Despite this, the regularization approach is useful for comparison, as it confirms two of our central findings with respect to the historical predictors. First, by far the most important predictor in the pre-2019 era is win status of the Director's Guild award for best director (that show's top honor), and second, since \textit{Roma} won that accolade in 2019, \textit{Roma} was well-supported as the historical favorite in this contest. Data and fully reproducible code for the analysis of the case study data is be available through the Journal's website.

\section{Discussion}
\label{sec:disc}

\subsection{Practical observations from journalist-statistician collaboration}

Nobody cares if you predict the Oscars after the awards have been announced.  An interesting aspect of this project that generalizes to many journalism settings is the comparatively short timeline available for analysis relative to other statistical consulting or academic settings.  From our experience, a typical start-to-finish timeline is about one-to-two weeks, spanning the conceptualization of the project, acquisition and organization of the data, data analysis and diagnostics, articulation of conclusions, and writing sufficient to appear in the appropriate venue.  Deadlines are hard, and falling behind leads to missed opportunities.

In the original \cite{Wilson2019we} analysis, there was a certain viable division of labor in the undertaking. Because data journalists invest a great deal of time learning to quickly and responsibly collect, format and fact-check datasets, we were able to start with a reasonably clean dataset of relevant awards. If all partners work in a common computing environment, it is possible to work off the same set of scripts aided by copious comments.

Yet, collaboration between academic statisticians and reporters who specialize in data-driven stories is not as common as it might be, given the wide disparity in deadlines and expectations. We aspire to offer some insight into how this gap can be bridged. Academic statisticians undergo a peer review process that is expected to take months for each article. These peer review articles are published at the discretion of the academic journal editor, who relies on feedback from associate editors and multiple reviewers.  By contrast, journalism editors typically assign an article ahead-of-time and are expecting the journalist to submit  text in anywhere from a few days to under an hour, depending on the breadth and depth of the story.

At the same time, the journalists who work primarily with quantitative sources--often called ``data journalists'' or ``computational journalists''--have neither the expertise nor the same burden of producing new research as academics. Still, given that, as established, a general-interest audience has a keen interest in predictions, any sort of partnership between our two fields benefits all parties by lending exposure to high-quality, innovative models and greatly enhancing the sophistication of the reporting.

That said, there are aspects of the original analysis for Time.com that were simplified in order to accommodate our rigid timeline. For example, the original analysis was based on standard logistic regression, which is unappealing since the predicted probabilities are in the context of the entire historical sample and do not sum to one within any given year. No subjective effects were incorporated in the original Time.com piece \citep{Wilson2019we}, because the methodological developments in Section 2 were not possible on the original timeline. This last point--the absence of any ability for readers to enter subjective effects, is an area of exciting continuing development. Long experience has demonstrated that interactivity, through the common Web tools like sliders and dropdowns, is a highly effective way to engage readers and attract a wide audience--not to mention giving readers the opportunity to explore the functionality of the underlying model. 

\subsection{Future directions}

In this work we have proposed the SB-PCLR statistical modeling approach that enables the researcher to obtain probability predictions for the winner of upcoming competitions based on historical data and, optionally, subjective inputs. The SB-PCLR method is valuable since it can be applied to any competition in which (a) historical data is available, (b) the number of winners and identities of participants in an upcoming round of the competition are known in advance, and (c) the method accommodates subjective input which can supplement historical effects when expert opinion can be elicited. We have demonstrated how the subjective Bayesian aspect of the work can be deployed in a broad readership with the online application accompanying this paper and in the subsequent 2020 Academy Awards \citep{Wilson2020our}.  The model can ``sit on top of '' other (not necessarily probabilistic) predictions by incorporating these into historical data where appropriate, or using them in the subjective specification aspect of the approach. For example, a user may incorporate pundit forecasts (data-driven or otherwise) into the subjective portion of their analysis even if those pundit forecasts do not extend back throughout the corpus of available historic data.  Thus, the SB-PCLR method would appear to be suitable for predicting the outcomes of award shows and sporting events.  The entertainment-based nature of these endeavors makes this a useful exercise as a public-facing opportunity for the readership in these areas to learn about subjective Bayesian approaches to data analysis.  Further, our online interactive interface system, when paired with an IRB-approved protocol, has  allowed for the collection of data which can be used to explore subjective elicitation in the readership audience for sports and movies. (We note that in our study, anonymity of participants was guaranteed and our IRB classified the project as non-human subjects research, but researchers must always involve an appropriate IRB when data on human subjects is of interest.) The vast majority of work on subjective elicitation focuses on studying \textit{expert} opinion \citep{Ohagan2006uncertain}, thus the information system the authors produced for the 2020 Time.com piece \citep{Wilson2020our} may provide the first look at large scale elicitation effects within our targeted readership. Obtaining the distribution of these effects is an important next step towards refining the uncertainty with which subjective probabilities and mixing weight $\omega$ described in Section \ref{sec:method} are expressed. While this manuscript shows the ability for MCMC to provide posterior estimates of uncertainty in the SB-PCLR framework, we have not yet deployed a method to obtain such intervals in real time for an online application for an upcoming competition. Our 2020 Application \citep{Wilson2020our} relied on MAP estimation.

Another potential avenue for future research would involve using and extending SB-PCLR methodology to study feedback loops between pundits and behavioral outcomes they predict. Forecasts from information age pundits may influence voters' opinions (in award shows or otherwise), thus creating a circular effect loop between pundits and decision makers. Pundits' forecasts influence decision makers, whose ultimate decisions get codified into historical data that pundits use for future forecasting efforts.

While much of the discussion in this paper has been focused on predicting upcoming competitions, the subjective machinery can be used \textit{post-mortem} to gauge e.g. the required degree of prior belief that would be necessary to ``move an incorrect historical prediction to the correct eventual winner'' in terms of probability. Beyond the entertainment domain of sport and film, the SB-PCLR method could potentially be useful for the study of elections, particularly given the still unresolved debate over why so few models or experts correctly predicted the 2016 U.S. presidential contest.

Since the method is Bayesian and we anticipate continuing to deploy our approach in web-enabled interfaces for a broad readership, this exercise provides an opportunity to educate the layperson about Bayesian methods.  The value in this is greater than advocating for a specific inferential paradigm, as the importance of incorporating human judgment in data-driven approaches becomes more salient by the day.  This method can be viewed as a sort of ``training ground'' for non-technical experts to grapple with these issues.  While the method relies on an extension of conditional logistic regression and hence grows out of the biostatistics tradition, outputs from black box machine learning (ML) algorithms can be incorporated into the subjective specification aspect of our approach, or the historical predictors where ML algorithms are applied to the entire corpus of historical data. Our SB-PCLR method can thus utilize historical data, subjective expertise, and modern prediction output to generate probability predictions for the outcome of upcoming competitions in a variety of human endeavors.

The SB-PCLR method does have drawbacks and opportunities for extension and improvement.  Model selection and averaging methodology could be further developed in this context, perhaps using recent results for automatic model selection-consistent mixture $g$ priors \citep[see e.g.][for an overview]{Li2018mixtures}, which would enable posterior summaries including credible intervals for regression coefficients and predictions in the context of model uncertainty. The ABMA approach presented here merely provides point estimates for these quantities. Further, the work here implicitly assumes the true model is in the candidate set. Recent work on the stacking of Bayesian predictive distributions has been shown to have favorable properties compared to Bayesian model averaging in the more realistic case where the true model is not in the candidate set \citep{Yao2018using}. Also, in settings where the number of competitors is large, forming the denominator of the conditional logistic regression likelihood in Equation (\ref{stratalike}) will be computationally expensive or unfeasible.

The SB-PCLR method assumes constant historical effects in all of the strata and independence among strata. Using the Academy Awards case study as an example, it seems unlikely that the $\bm{\beta}$ effects are truly constant in time. Future research in time-varying coefficients within the SB-PCLR framework would enable a diagnosis of this possibility. Such an approach would also address the implicit equal-weighting given to all strata, where perhaps next year's prediction should rely more heavily on recent effects compared to more remote historical effects. Further, it seems likely that in contiguous years many voters are the same. The effect of these voters' dispositions is not recognized by the model, which instead treats data from adjoining strata as independent.

\bibliographystyle{chicago}
\bibliography{refs}

\section*{Acknowledgments}
The authors would like to thank the following organizations: The Internet Movie Database,	Oscars.org, DGA.org, BAFTA.org, GoldenGlobes.com, and OMDBapi.com.

\newgeometry{bottom=1cm}
\begin{landscape}
\begin{table}[ht]
\caption{Predicted win probabilities for the 2019 Academy Award nominees for Best Picture. Second column is based on standard (non-conditional) logistic regression. Third column is the posterior predicted probability of winning with no subjective effects.  Columns four-six, seven-nine, and ten-twelve, respectively  include posterior probabilities at $\omega=0.5$ and $\omega =0.2$, and the prior probabilities (i.e. $\omega=0$) for each setting. Final column includes win probabilities based on approximate Bayesian model averaging via BIC.}
\label{Table1}
\centering
\begin{tabular}{|l|c|c|c|ccc|ccc|ccc|c|}
  \hline
  & & & & \multicolumn{3}{c|}{GB prior}& \multicolumn{3}{c|}{Um prior}& \multicolumn{3}{c|}{NR prior} &  \\
Film name & Logistic & LASSO & $\omega=1.0$ & $\omega=0.5$ & $\omega=0.2$ & prior & $\omega=0.5$ & $\omega=0.2$ & prior & $\omega=0.5$ & $\omega=0.2$ & prior & ABMA\\ 
  \hline
A Star Is Born & 0.00 & 0.07 & 0.00 & 0.02 & 0.02 & 0.03 & 0.06 & 0.10 & 0.12 & 0.07 & 0.11 & 0.14 & 0.02\\ 
  Black Panther & 0.00 & 0.07 & 0.00 & 0.02 & 0.02 & 0.03 & 0.06 & 0.10 & 0.12 & 0.07 & 0.11 & 0.14 & 0.01\\ 
  BlacKkKlansman & 0.11 & 0.08 & 0.20 & 0.11 & 0.06 & 0.03 & 0.16 & 0.14 & 0.12 & 0.17 & 0.15 & 0.14 & 0.17\\ 
  Bohemian Rhapsody & 0.02 & 0.09 & 0.03 & 0.03 & 0.03 & 0.03 & 0.08 & 0.11 & 0.12 & 0.09 & 0.12 & 0.14 & 0.04\\ 
  Green Book & 0.02 & 0.08 & 0.03 & 0.41 & 0.65 & 0.80 & 0.08 & 0.11 & 0.12 & 0.09 & 0.12 & 0.14 & 0.12\\ 
  Roma & 0.46 & 0.60 & 0.34 & 0.19 & 0.09 & 0.03 & 0.23 & 0.17 & 0.12 & 0.18 & 0.08 & 0.01 & 0.40\\ 
  The Favourite& 0.11 & 0.08 & 0.20 & 0.11 & 0.06 & 0.03 & 0.16 & 0.14 & 0.12 & 0.17 & 0.15 & 0.14 & 0.16\\ 
  Vice & 0.11 & 0.08 & 0.20 & 0.11 & 0.06 & 0.03 & 0.16 & 0.14 & 0.12 & 0.17 & 0.15 & 0.14 & 0.07\\ 
   \hline
\end{tabular}
\end{table}
\end{landscape}
\restoregeometry

\begin{table}[ht]
\caption{Predicted win probabilities and 95 \% equal-tail credible intervals based on MCMC sampler.}
\label{Table2}
\centering
\begin{tabular}{lllll}
  \hline
   & & \multicolumn{3}{c}{GB prior}   \\
Film name & $\omega=1.0$ & $\omega=0.5$ & $\omega=0.2$ & $\omega=0$ (i.e. prior)\\ 
  \hline
A Star Is Born & 0.00(0.00,0.02) & 0.02(0.01,0.02) & 0.02(0.02,0.03) & 0.029 \\ 
  Black Panther & 0.00(0.00,0.02) & 0.02(0.01,0.02) & 0.02(0.02,0.03) & 0.029 \\ 
  BlacKkKlansman & 0.19(0.11,0.26) & 0.11(0.07,0.15) & 0.06(0.05,0.08) & 0.029 \\ 
  Bohemian Rhapsody & 0.03(0.00,0.09) & 0.03(0.02,0.06) & 0.03(0.02,0.04) & 0.029 \\ 
  Green Book & 0.03(0.00,0.09) & 0.42(0.40,0.45) & 0.65(0.64,0.66) & 0.800 \\ 
  Roma & 0.35(0.15,0.60) & 0.19(0.09,0.32) & 0.09(0.05,0.14) & 0.029 \\ 
  The Favourite & 0.19(0.11,0.26) & 0.11(0.07,0.15) & 0.06(0.05,0.08) & 0.029 \\ 
  Vice & 0.19(0.11,0.26) & 0.11(0.07,0.15) & 0.06(0.05,0.08) & 0.029 \\ 
   \hline
\end{tabular}
\end{table}

\begin{table}[ht]
\centering
\begin{tabular}{lllll}
  \hline
   & & \multicolumn{3}{c}{U prior}   \\
Film name & $\omega=1.0$ & $\omega=0.5$ & $\omega=0.2$ & $\omega=0$ (i.e. prior) \\ 
  \hline
A Star Is Born & 0.00(0.00,0.02) & 0.06(0.06,0.07) & 0.10(0.10,0.10) & 0.125 \\ 
  Black Panther & 0.00(0.00,0.02) & 0.06(0.06,0.07) & 0.10(0.10,0.10) & 0.125 \\ 
  BlacKkKlansman & 0.19(0.11,0.26) & 0.16(0.12,0.19) & 0.14(0.12,0.15) & 0.125 \\ 
  Bohemian Rhapsody & 0.03(0.00,0.09) & 0.08(0.06,0.11) & 0.11(0.10,0.12) & 0.125 \\ 
  Green Book & 0.03(0.00,0.09) & 0.08(0.06,0.11) & 0.11(0.10,0.12) & 0.125 \\ 
  Roma & 0.35(0.15,0.60) & 0.24(0.14,0.36) & 0.17(0.13,0.22) & 0.125 \\ 
  The Favourite & 0.19(0.11,0.26) & 0.16(0.12,0.19) & 0.14(0.12,0.15) & 0.125 \\ 
  Vice & 0.19(0.11,0.26) & 0.16(0.12,0.19) & 0.14(0.12,0.15) & 0.125 \\ 
   \hline
\end{tabular}
\end{table}

\begin{table}[ht]
\centering
\begin{tabular}{lllll}
  \hline
   & & \multicolumn{3}{c}{NR prior}   \\
Film name & $\omega=1.0$ & $\omega=0.5$ & $\omega=0.2$ & $\omega=0$ (i.e. prior) \\ 
  \hline
A Star Is Born & 0.00(0.00,0.02) & 0.07(0.07,0.08) & 0.11(0.11,0.12) & 0.141 \\ 
  Black Panther & 0.00(0.00,0.02) & 0.07(0.07,0.08) & 0.11(0.11,0.12) & 0.141 \\ 
  BlacKkKlansman & 0.19(0.11,0.26) & 0.17(0.13,0.20) & 0.15(0.14,0.17) & 0.141 \\ 
  Bohemian Rhapsody & 0.03(0.00,0.09) & 0.09(0.07,0.12) & 0.12(0.11,0.13) & 0.141 \\ 
  Green Book & 0.03(0.00,0.09) & 0.09(0.07,0.12) & 0.12(0.11,0.13) & 0.141 \\ 
  Roma & 0.35(0.15,0.60) & 0.18(0.08,0.31) & 0.08(0.04,0.13) & 0.010 \\ 
  The Favourite & 0.19(0.11,0.26) & 0.17(0.13,0.20) & 0.15(0.14,0.17) & 0.141 \\ 
  Vice & 0.19(0.11,0.26) & 0.17(0.13,0.20) & 0.15(0.14,0.17) & 0.141 \\ 
   \hline
\end{tabular}
\end{table}

\newgeometry{bottom=1cm}
\begin{landscape}
\begin{table}[ht]
\caption{Summary of top 10 candidate predictors as determined by odds ratio of winning Best Picture. Posterior mean and 95\% equal tail credible intervals presented for coefficients based on univariate regression and the selected model using the SB-PCLR method. LASSO and ABMA coefficients are in the final two columns. Dash entries `-' correspond to variables that have coefficients of zero. }
\label{Table3}
\centering
\begin{tabular}{lccccc}
  \hline
Predictor name & Odds ratio & Univariate $\beta_j$ & Model $\beta_j$ & LASSO & ABMA\\ 
  \hline
Won best director - Director's Guild of America & 66.60 & 2.78(2.22,3.4) & 2.54(1.89,3.28) & 2.96 & 2.57\\ 
  Nominated for best director - Director's Guild of America & 30.22 & 3.65(1.85,6.5) & - & - & 0.12 \\ 
  Nominated for best director - Academy Awards & 20.66 & 3.07(1.79,4.85) & 2.06(0.51,3.93) & - & 0.50\\ 
  Nominated for best director - Golden Globes & 10.28 & 2.3(1.4,3.38) & - & - & 0.01\\ 
  Nominated for best film editing - Academy Awards & 8.63 & 2.03(1.29,2.92) & 1.99(0.99,3.16) & 0.10 & 1.79 \\ 
  Won best picture in drama - Golden Globes & 8.02 & 1.51(1.03,2.01) & - & 0.13 & 0.16\\ 
  Won best director - Golden Globes & 7.84 & 1.52(1.03,2.03) & - & - & -0.04\\ 
  Won best film - British Academy of Film and Television Arts & 5.68 & 1.54(0.96,2.17) & - & - & 0.05\\ 
  Won best cinematography - British Academy of Film and Television Arts & 5.50 & 1.61(0.87,2.35) & - & - & 0.06\\ 
  Nominated for best film - British Academy of Film and Television Arts & 4.60 & 1.76(1.1,2.49) & - & - & 1.09\\ 
   \hline
\end{tabular}
\end{table}
\end{landscape}
\restoregeometry
\end{document}